# MegaOhm Extraordinary Hall effect in oxidized CoFeB.

G. Kopnov and A. Gerber

Raymond and Beverly Sackler Faculty of Exact Sciences

School of Physics and Astronomy

Tel Aviv University

Ramat Aviv, 69978 Tel Aviv, Israel

We report on development of controllably oxidized CoFeB ferromagnetic films demonstrating the extraordinary Hall effect (EHE) resistivity exceeding 1 $\Omega$cm and magnetic field sensitivity up to $10^6$ $\Omega$/T. Such EHE resistivity is four orders of magnitude higher than previously observed in ferromagnetic materials, while sensitivity is two orders larger than the best of semiconductors.

Spin Hall effect and its implementation in ferromagnetic materials, known as the extraordinary or anomalous Hall effect (EHE), attracts significant interest in recent years. Physical mechanisms of the phenomena are under scrutiny [1,2] as well as it's possible applications for sensors, memories and logic devices [3,4]. Remarkably high magnetic field sensitivity beyond $10^4$ Ω/T, which is larger than the best Hall effect sensitivity ever found in semiconducting materials [5], was recently reported [6,7]. The strategy adapted to achieve such sensitivity was by using very thin ferromagnetic films with enhanced spin-orbit surface scattering and tailored perpendicular magnetic anisotropy that enabled easy out-of-plane rotation of magnetization with low saturation field [8]. While stand-alone EHE sensors are ready for use, their compatibility with future generation of CMOS technology was critically analyzed [4]. It was estimated that in order to couple an EHE cell to FET (field effect transistor) in microelectronic circuits with 22 nm line width, resistivity of the material should exceed 50 mΩcm, while EHE resistivity to be larger than 2.5 mΩcm. The highest room temperature EHE resistivity of about 200 μΩcm, named the Giant Hall effect [9], was found in granular ferromagnetic mixtures, such as Ni-$SiO_2$ [9,10] and CoPt-$SiO_2$ [4], in vicinity of the conductance percolation threshold. It was therefore concluded [4] that while EHE can readily be used for micrometer and larger devices, no materials are yet available which offer suitable scalability towards the nanometer size microelectronic node.

In this letter, we report on development of low conductivity ferromagnetic material demonstrating controllable resistivity up to $10^2$ Ωcm, EHE resistivity exceeding 1 Ωcm and field sensitivity up to $10^6$ Ω/T. Such EHE resistivity is four orders of magnitude higher than previously observed in ferromagnetic materials, while the sensitivity is two orders larger than the finest of semiconductors.

Two methods are known to effectively increase resistivity and EHE in ferromagnetic metals: reducing film thickness that leads to enhancement of surface spin-orbit scattering [11]; and mixing with immiscible insulator materials, like $SiO_2$, leading to a similar effect at intergranular metal-insulator interfaces [9]. Polycrystalline metallic ferromagnetic films (Ni, Co, Fe) fabricated by conventional vacuum deposition methods, such as sputtering and e-beam deposition, become electrically disconnected when their averaged thickness is reduced below 1-3 nm. The highest room temperature resistivity typically observed in such films is 0.1 – 1 mΩcm, and EHE resistivity is 1–10 μΩcm [11]. Three-dimensional granular films, composed of immiscible metallic ferromagnet and insulator mixtures, reach geometrical percolation threshold when the insulator volume content exceeds 40-50% (80% in random metal-insulator mixtures) with characteristic values of longitudinal and EHE resistivities of 1 Ωcm and 100 μΩcm respectively [9]. In this work we explored a different approach to fabrication of low conductance films: we start with amorphous ferromagnetic metal and increase its resistivity by gradual oxidation. The selected material is CoFeB ($Co_{40}Fe_{40}B_{20}$). Series of thin films with different thickness (5 nm $\leq t \leq$ 100 nm) were fabricated by RF magnetron sputtering from $Co_{40}Fe_{40}B_{20}$ target (ACI Alloys Inc.) on rectangular 5x5mm$^2$ pieces of intrinsic GaAs substrate. Base pressure prior to deposition was about $2\times10^{-7}$ torr, whereas deposition took place at $5\times10^{-3}$ torr Ar atmosphere mixed with controlled flow of either air or pure oxygen. Typical deposition rate was 0.1-0.2 nm/sec. Resistance, magnetoresistance and Hall effect were measured using Van der Pauw protocol. The data shown here, with exception of Fig.5a, were measured at room temperature.

Effect of reactive sputtering on resistivity is shown in Fig.1 for a series of 40 nm thick samples deposited under variable air flow. Films deposited in a pure Ar (99.995%) atmosphere had typical resistivity in $10^{-4}$ Ωcm range. Minor addition of air increased resistivity to $10^{-3} – 10^{-2}$ Ωcm. Resistivity grows sharply up to $10^2$ Ωcm when partial air pressure goes above $10^{-3}$ torr, that is about 1:5 ratio with argon. XRD (Fig.1 inset) and TEM images of films sputtered in pure Ar and in Ar/air atmosphere are typical for amorphous films. One should note that films exposed to electron beam for an extended period of time start to show local crystallization, which is consistent with a known effect of crystallization under post-deposition annealing [12]. Resistivity of unprotected samples increases with time, probably due to additional oxidation.

To remind, Hall resistance $R_{xy}$ in magnetic films can be presented as:

$$R_{xy} = V_{xy}/I = \rho_{xy}/t = (R_0 B_z + \mu_0 R_s M_z)/t,$$

where $\rho_{xy}$ is Hall resistivity, $I$ – current, $t$ - thickness, $R_0$ and $R_s$ are the ordinary and the extraordinary Hall effect coefficients and $B_z$ and $M_z$ are the normal-to-plane projections of magnetic field induction and magnetization, respectively [13]. EHE resistivity $\rho_{EHE}$, discussed in the following, is defined as: $\rho_{EHE} = \mu_0 R_s M_{z,sat}$, where $M_{z,sat}$ is the saturated out-of-plane magnetization, and is determined by extrapolation of the magnetically saturated high field linear portion of $R_{xy}(B)$ to zero field.

Fig.2 presents the absolute value of Hall resistance $R_{xy}$ as a function of magnetic field applied perpendicular to film plane for several samples deposited at different partial air pressures. Note that $R_{xy}$ scale is logarithmic. Sample 1 was deposited in a pure Ar atmosphere; its saturated EHE resistance is about 1Ω which is typical for thin ferromagnetic films. The signal grows dramatically in oxidized samples, reaching unprecedented $10^6$ Ω in sample 4. We measured a large number of samples of different thickness fabricated at different oxygen/air flow ratios, at slightly different deposition rates, at various starting base pressures, and at different aging states after the deposition. For all samples we found a very good and reproducible correlation between the EHE resistivity $\rho_{EHE}$ and the longitudinal resistivity $\rho$ shown in Fig.3. Dramatic changes in $\rho_{EHE}$ occur when resistivity exceeds 0.1 Ωcm: (i) magnitude of the effect grows sharply as $\rho_{EHE} \propto \rho^2$ and exceeds 1 Ωcm; (ii) polarity of the effect reverses from negative in low resistivity samples to positive in high resistivity ones. Notably, polarity of the ordinary Hall effect measured at high fields beyond magnetic saturation remains negative in all samples studied.

An important parameter characterizing magnetic field sensors is sensitivity, defined as: $S = dR_{xy}/dB$. Fig.4 presents the absolute value of sensitivity as a function of sheet resistance for thin films. Consistently with the data shown in Fig.3, sensitivity starts growing dramatically when sheet resistance exceeds 10 kΩ/□ threshold, reaching very high values of order $10^6$ Ω/T. As mentioned above, high EHE sensitivity of order $10^4$ Ω/T reported in Refs. [6,7] was achieved in ultrathin films with $\rho_{EHE}$ of few μΩcm by reducing the saturation field to $10^{-3}$ T. Our samples have no out-of-plane anisotropy, saturation field remains high (about 1.5 T), and high sensitivity is achieved due to very large EHE resistivity $\rho_{EHE}$.

An obvious question rises, what makes the extraordinary Hall effect in oxidized CoFeB so different from previously studied systems? Judging by transmission electron microscopy and x-ray diffraction the material is amorphous. Resistivity temperature dependence of oxidized samples, shown in Fig.5a, follows the variable range hopping model, as: $\rho = \rho_0 exp\left(\frac{T_0}{T}\right)^{1/4}$ with $T_0$ increasing from 1 K for sample 1 with room temperature resistivity $10^{-2}$ Ωcm to 25 K for sample 3 with resistivity 3.2 Ωcm. Such temperature dependence was observed in many disordered amorphous and granular systems below percolation threshold, including granular $NiSiO_2$ mixtures [10]. Field dependence of EHE resistivity $\rho_{EHE}(B)$ (Fig. 5b) can be well fitted by Langevin function $L(x) = \coth(x) - 1/x$, with $x = MB/k_BT$. The behavior is typical for paramagnetic and superparamagnetic systems above blocking temperature. Magnetic moment $M$ extracted from the fitting is about 1.2x$10^{-20}$ A/m. Assuming that magnetization of magnetic clusters is that of bulk $Co_{40}Fe_{40}B_{20}$ [14], we estimate their diameter as about 3 nm. One can then visualize the material as composed of weakly coupled magnetic clusters with electrical conductance governed by variable range hopping or thermally activated tunneling. Magnetoresistance measurements, shown in Fig. 5c, support this scheme. Two data sets were obtained with magnetic field applied in-plane of the sample perpendicular and parallel to current direction. Magnetoresistance is negative and isotropic, which is characteristic for spin-dependent tunneling magnetoresistance in granular ferromagnets below percolation threshold [15] and in weakly coupled thin ferromagnetic films [16]. Thus, apart from the extraordinary Hall effect, magnetotransport properties of oxidized CoFeB are qualitatively similar to those observed in other disordered ferromagnetic materials.

Modern models of the extraordinary Hall effect predict three conductance ranges with qualitatively different correlations between longitudinal resistivity and the EHE [17,18]. In the clean regime ($\rho < 10^{-6}$ Ω$cm$), the skew scattering mechanism, for which EHE resistivity scales linearly with resistivity ($\rho_{EHE} \propto \rho$), is predicted to dominate. The intrinsic EHE mechanism is predicted to dominate in the intermediate disorder regime ($\rho \sim 10^{-6} - 10^{-4}$ Ωcm) with quadratic scaling ($\rho_{EHE} \propto \rho^2$). In the high disorder range ($\rho > 10^{-4}$ Ω$cm$) the intrinsic contribution is strongly decayed, resulting in a scaling relation $\rho_{EHE} \propto \rho^\gamma$ with $\gamma \sim 0.4$. This theory is based on the use of Bloch wave functions assuming a metallic conduction, thus in principle the result is valid only for ferromagnetic metals. On the other hand, similar scaling $\rho_{EHE} \propto \rho^\gamma$ with $0.24 \leq \gamma \leq 0.67$ has also been predicted [19] for thermally activated hopping processes like variable range hopping, short-range activation hopping or tunneling influenced by interactions in the Efros-Shklovskii regime. Universal scaling in the form $\rho_{EHE} \propto \rho^\gamma$ with $\gamma \sim 0.4$ is therefore anticipated for low conductivity materials regardless whether their conductivity is metallic or thermally activated. We are not aware of any alternative model predicting power index $\gamma = 2$ in the high resistivity limit.

Change of EHE polarity is another intriguing phenomenon. Variation of EHE polarity with composition was observed in metallic ferromagnetic alloys NiFe [20], TbCo [21] and CoPd [22, 23]. Split-band model was used by L. Berger [20] to interpret the effect in NiFe, but it is questionable whether this model can be relevant in our case.

To summarize, partially oxidized CoFeB films exhibit huge extraordinary Hall effect resistivity exceeding 1 Ωcm and field sensitivity up to $10^6$ Ω/T. This EHE resistivity is four orders of magnitude higher than previously observed in ferromagnetic materials, while sensitivity is two orders larger than the best of semiconductors. The outstanding properties: magnitude of the effect, change of polarity and quadratic power law scaling of the EHE resistivity with the longitudinal one in the high resistivity limit are puzzling and need to be explored and understood.

This research was supported in part by TAU Nano-Center grant No.060328561.

**References.**

**Figure Captions.**

Fig.1. Resistivity of a series of RF sputtered CoFeB films as a function of partial air pressure in Ar/air atmosphere. Ar pressure is 5×10$^{-3}$ torr. Films are 40 nm thick. Inset: typical XRD image detected for non-oxidized and oxidized samples.

Fig.2. Hall effect resistance as a function of magnetic field applied normal to the film plane for a number of samples deposited under increasing partial air pressures. Data for sample 1 are multiplied by (-1). Thickness and deposition air pressure for the presented samples are respectively: 1 - 30 nm, 0 mtorr; 2 – 10 nm, 1.25 mtorr; 3 – 10 nm, 1.3 mtorr; 4 – 25 nm, 1.6 mtorr. Error bars in this and following figures are smaller than the symbols size.

Fig.3. Absolute values of the EHE resistivity as a function of longitudinal resistivity. Vertical line indicates transition between samples with negative and positive $\rho_{EHE}$.

Fig.4. Hall effect sensitivity (absolute values) as a function of sheet resistance for a variety of thin samples.

Fig.5. Magnetotransport characterization of partially oxidized CoFeB films. (a) resistivity as a function of temperature ($T^{-1/4}$) for three samples with room temperature resistivity 1.02×10$^{-2}$ Ωcm (1), 1.4×10$^{-2}$ Ωcm (2) and 3.2 Ωcm (3). Left vertical axis corresponds to sample 3. Solid lines are linear fits. (b) EHE resistance as a function of field applied normal to the film plane (symbols). The ordinary Hall effect contribution was subtracted. Solid line is a fit to Langevin function. (c) Normalized magnetoresistance measured under in-plane magnetic field applied perpendicular (○) and parallel (●) to current. $\rho \approx 7$ Ωcm.

**Figures.**

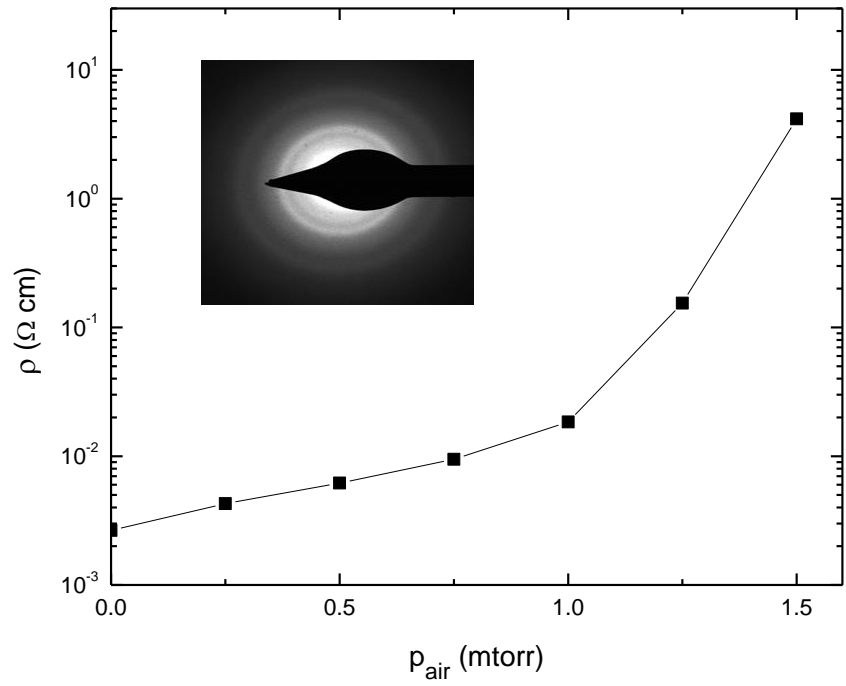

Fig. 1

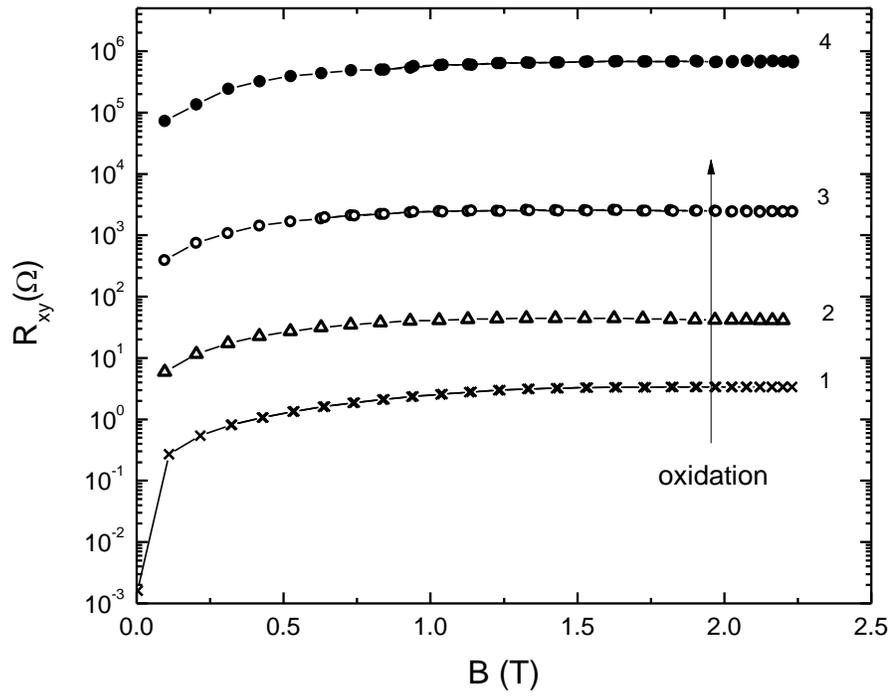

Fig. 2

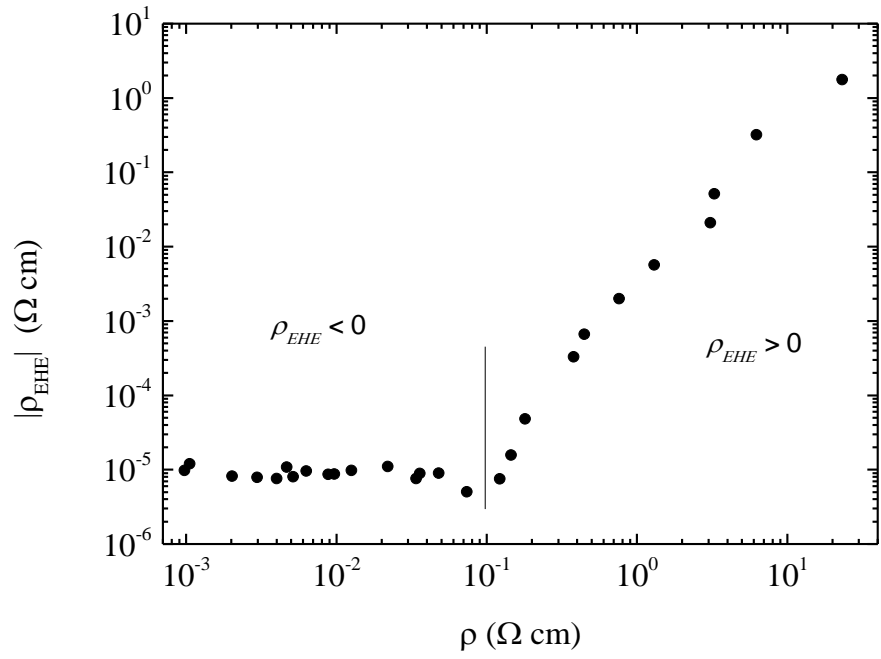

Fig. 3

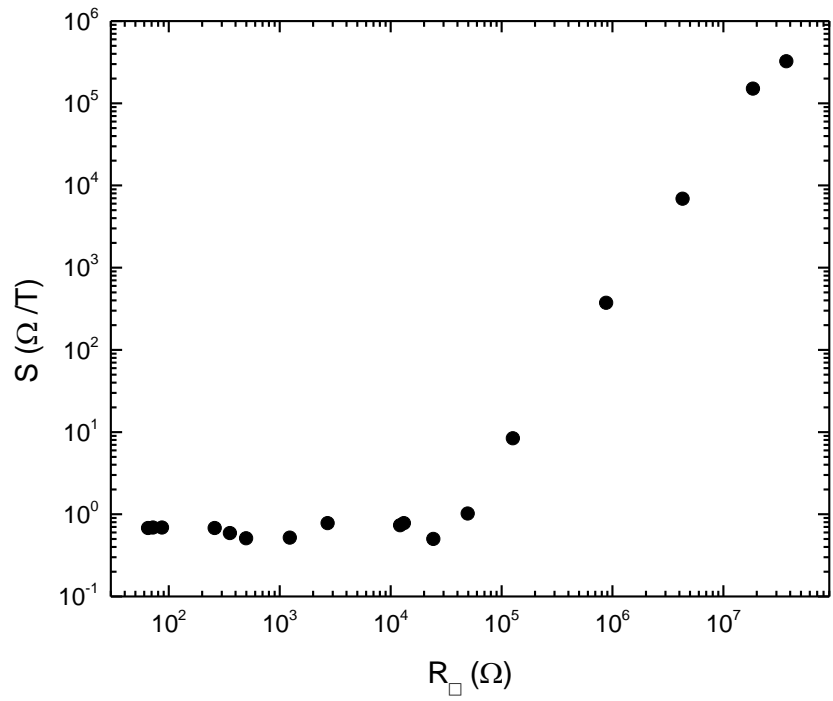

Fig. 4

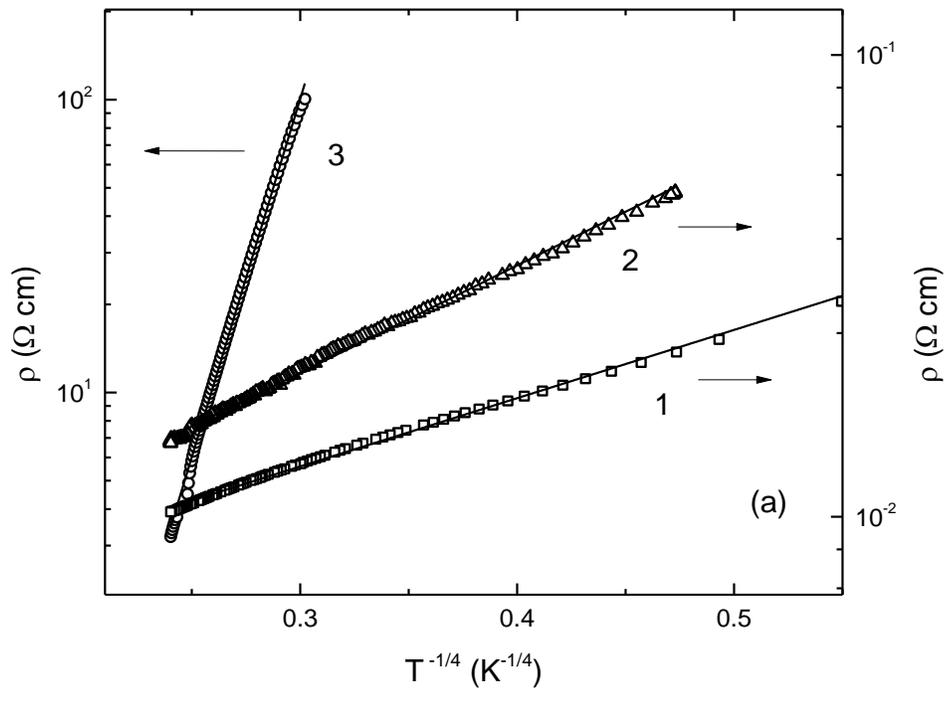

Fig. 5a

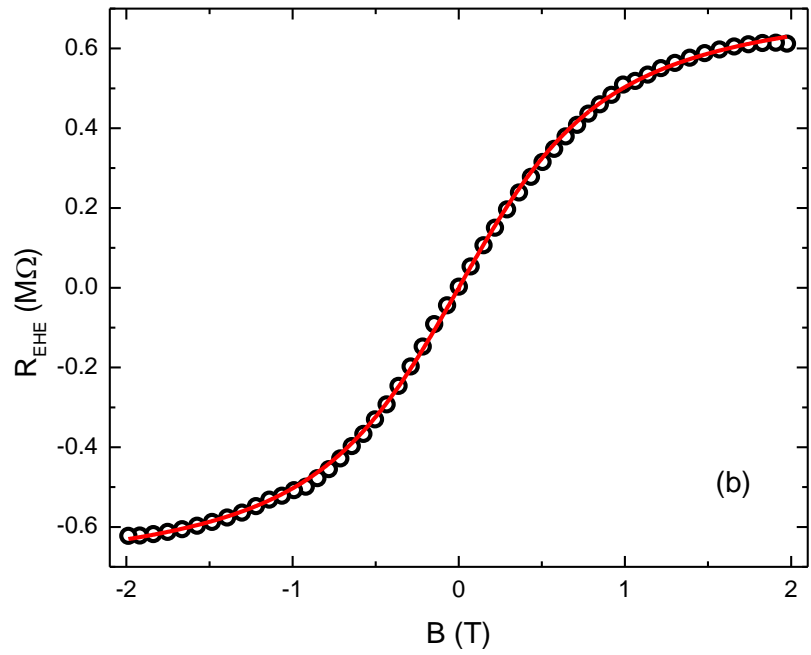

Fig. 5b

Fig. 5c